# Emphasizing Cherenkov photons from bismuth germanate by single photon deconvolution

Ryosuke Ota, *Member, IEEE*, and Kibo Ote

*Abstract*—Bismuth germanate (BGO) has been receiving attention again because it is a potential scintillator for future time-of-flight positron emission tomography. Owing to its optical properties, BGO emits a relatively large number of Cherenkov photons after 511 keV gamma-ray interactions, which pushes the timing resolution of a detector. Nonetheless, efficiently detecting Cherenkov photons among scintillation photons is similar to looking for a needle in a haystack. Thus, we propose a method that can efficiently emphasize Cherenkov photon from a detector waveform by deconvolving a single photon response of photodetector. As a proof-of-concept, we perform the deconvolution, and a probability density function (PDF) of bismuth germanate was obtained, which is compared to a conventional time correlated single photon counting method. Furthermore, we investigate if the proposed deconvolution can emphasize a faint Cherenkov photon. Consequently, the PDF obtained by the proposed deconvolution shows a good agreement with that obtained using a conventional method. A coincidence time resolution obtained using the proposed deconvolution is improved by 43% in full width at half maximum, compared to a voltage-based leading edge discriminator. It can be concluded that the proposed deconvolution method can efficiently emphasize Cherenkov photon and improve the timing performance of BGO-based detectors.

## I. INTRODUCTION

BISMUTH germanate (BGO) is one of the scintillators most commonly used for radiation measurements. In the field of positron emission tomography (PET), BGO was the main scintillator until the 1990s [1]. Since the emergence of scintillators that are faster than BGO, such as lutetium oxyorthosilicate (LSO:Ce) [2] and lutetium-yttrium oxyorthosilicate (LYSO) [3], researchers have focused on their use in time-of-flight (TOF) PET. Furthermore, in the 2000s, silicon photomultipliers (SiPMs) became commercially available and are currently unrivaled photodetectors for TOF-PET, but not limited to it, owing to their excellent photon detection efficiency (PDE) and single photon time resolution (SPTR) compared with conventional photomultiplier tubes (PMT) [4,5]. These technological advancements have made TOF-PET systems commercially available with system coincidence time resolutions (CTRs) of 200-400 ps full width at half maximum (FWHM) [6-8]. At the laboratory level, an SiPM readout using high-frequency readout electronics achieves an SPTR of 70 ps FWHM while having a high PDE of 60% at 410 nm, reaching a CTR of 58 ps FWHM when coupled with a 2×2×3 mm³ calcium co-doped LSO:Ce [9-11]. Thus, the combination of SiPM and lutetium-based scintillator is the mainstream development of TOF-PET detectors. However, two recent studies triggered the reuse of BGOs for TOF-PET detectors by detecting Cherenkov photons promptly produced in BGO; consequently, BGO has received the attention of researchers in the past few years [12, 13].

Cherenkov photons are emitted when a relativistic charged particle passes through a dielectric medium. Cherenkov photons convey timing information more accurately than scintillation photons because they are promptly emitted on the order of picoseconds (ps), achieving a CTR of ~30 ps FWHM when combined with ultrafast photodetectors and machine learning [14-16]. In contrast, the number of emitted Cherenkov photons was approximately three orders of magnitude lower than that of the scintillation photons. The expected number of emitted photons $N$ can be described as follows:

$$N \propto 2\pi\alpha L \left(\frac{1}{\lambda_1} - \frac{1}{\lambda_2}\right)\left(1 - \frac{1}{(n\beta)^2}\right), \quad (1)$$

where $\alpha$, $L$, $\lambda$, $n$, and $\beta$ are the fine structure constant, path length of a charged particle, wavelength of emitted photons, refractive index of the dielectric medium, and velocity of the charged particle relative to the speed of light, respectively. Here, $\lambda_1 < \lambda_2$ and practically, $\lambda_1$ represents the cutoff wavelength of the dielectric material. Therefore, the higher the refractive index, the larger the number of emitted photons. BGO is a good candidate in terms of Cherenkov emission and detection because not only does it have a high refractive index of ~2.2 at 400 nm [17] and transparent down to 300 nm (meaning $\lambda_1$ can be set to 300 nm) but also it has relative slower decay time constants of 45.8 and 365 ns [18], facilitating Cherenkov/scintillation separation. Nonetheless, Cherenkov/scintillation separation remains challenging but should be addressed to fully extract the potential of BGO by detecting the first photon [19].

*Corresponding author: Ryosuke Ota*

Ryosuke Ota and Kibo Ote are with the Central Research Laboratory, Hamamatsu Photonics K.K., Hamamatsu, Japan (e-mail: ryosuke.ota@crl.hpk.co.jp).

The dual-ended readout scheme is an efficient method for identifying Cherenkov- and scintillation-triggered events [20]. By alternatively selecting a faster SiPM signal, scintillation-triggered events were efficiently suppressed, thereby reducing the long-tail of the coincidence histogram. Another technique utilizes the rise time information of a detector signal, which varies depending on the number of detected Cherenkov photons [21,22]. These techniques benefit from high-frequency readout electronics [10] because they have a high amplitude for a single photon with a wide bandwidth, emphasizing the first detected photon signal. However, these techniques do not retrieve all the detected photon information, which could be useful for better extraction of Cherenkov-triggered events.

In this study, we propose an efficient method for extracting Cherenkov photons by deconvolving the single photon response of a multipixel photon counter (MPPC) from a detector signal on an even-by-event basis. By deconvolution, we can retrieve all the detected photon information (i.e., the probability density function (PDF) of the BGO emission combined with the photon travelling time in the crystal), including the Cherenkov photons, even if the Cherenkov signal is buried in electric noise to some extent. Although a deconvolution method of a single photon response from a detector signal has already been proposed in [23], a PMT with a single photon response in terms of the amplitude that largely varies event-by-event compared to the MPPC, was used as a photodetector. Therefore, using the MPPC is more robust; thus, we can expect to acquire a more accurate PDF. In this study, we demonstrate the accuracy of the proposed deconvolution method by comparing it with a conventional time correlated single photon counting method (TCSPC) and demonstrate that it has a better CTR than the conventional voltage-based leading edge discriminator CTR.

## II. MATERIALS AND METHODS

### A. Concept of the proposed method

Waveforms $w(t)$ from a detector consisting of a scintillator and a photodetector are formed by convolving the scintillation kinetics $f(t)$ and single photon response $i(t)$ and the SPTR $s(t)$ of the photodetector, and the addition of electric noise, as illustrated in Fig. 1. In this study, we chose BGO ($3 \times 3 \times 15$ mm$^3$, all polished) and MPPC (S13360-3050CS) as the scintillator and photodetector, respectively; the terms $f_{BGO}(t)$, $i_{MPPC}(t)$, $s_{MPPC}(t)$ are used, and the waveforms are described as follows:

$$w(t) = f_{BGO}(t) * i_{MPPC}(t) * s_{MPPC}(t) + noise, \quad (2)$$

where $*$ denotes the convolutional operator. If noises are considered negligible, $f_{BGO}(t)$ can be calculated on an even-by-event basis by $w(t) *^{-1} (i * s)_{MPPC}(t)$, where $*^{-1}$ and $(i * s)_{MPPC}(t)$ are the deconvolution operator and $i_{MPPC}(t) * s_{MPPC}(t)$, respectively, because $i_{MPPC}(t)$ can be considered constant unlikely to $i_{PMT}(t)$. Given that the intrinsic SPTR of the MPPC is ~150 ps FWHM [11], it can be expected to extract Cherenkov photons from the rinsing edge of the waveform, as reported in [13, 18], as long as the single photon amplitude is higher than the noise floor. The probabilities of crosstalk after an MPPC pulse are discussed later.

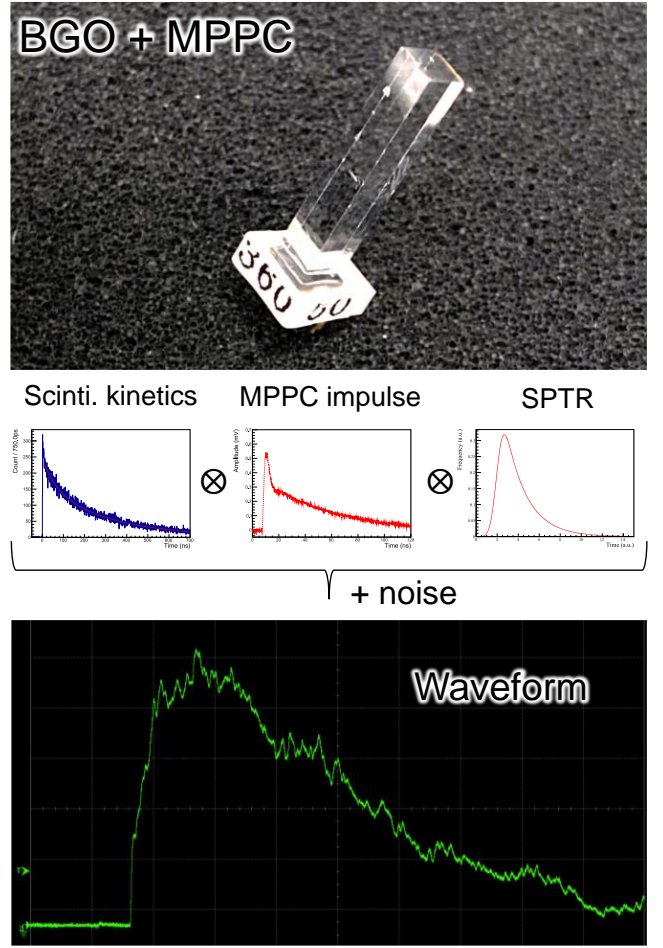

**Fig. 1.** Concept of this study. Waveforms of the detector consisting of a BGO and an MPPC can be described by convolving the BGO scintillation kinetics $f_{BGO}(t)$, the single photon response $i_{MPPC}(t)$ and SPTR $s_{MPPC}(t)$. This indicates that $f_{BGO}(t)$ can be obtained event-by-event by deconvolving $i_{MPPC}(t) * s_{MPPC}(t)$ from the waveform

### B. Measurement of the single photon response

To perform deconvolution, $(i * s)_{MPPC}(t)$ should be measured in advance. Fig. 2(a) shows the experimental setup used for the measurement. A pulse laser with 60 ps FWHM pulse width (C10196, Hamamatsu Photonics K.K.) irradiates the MPPC with a neutral density filter to decrease the laser intensity down to single photon level. A voltage of 60.16 V was supplied to the MPPC, which equals to the overvoltage from the breakdown voltage (V$_{ov}$) of 8.0 V to make the single photon amplitude higher than the electric noise. Signals from the MPPC were fed into an oscilloscope (DSOS404A, KEYSIGHT) with a bandwidth and sampling rate of 4.2 GHz and 20 GS/s, respectively, without using an amplifier, and waveforms were recorded event-by-event. The vertical range of the oscilloscope was set to 8 mV (1 mV/div). with 10 bit ADC. The oscilloscope was triggered by a synchronized laser signal to ensure the triggering time of the MPPC signal. The dark signals of the MPPC may also be useful for single photon response measurements; however, the trigger time cannot be assured because of the small amplitude of the MPPC relative to the electric noise. Therefore, we decided to use a pulsed laser. The maximum amplitude of the recorded waveform was extracted on an event-by-event basis and filled into a histogram, as shown in Fig. 2(b). After filling all the events, single photon equivalent events within $\pm 1\sigma$ (red shaded area in the Fig. 2(b)) were accepted, and finally, the waveforms of the accepted events were averaged with baseline correction as depicted in Fig. 2(c). The number of events used for averaging was 1,119.

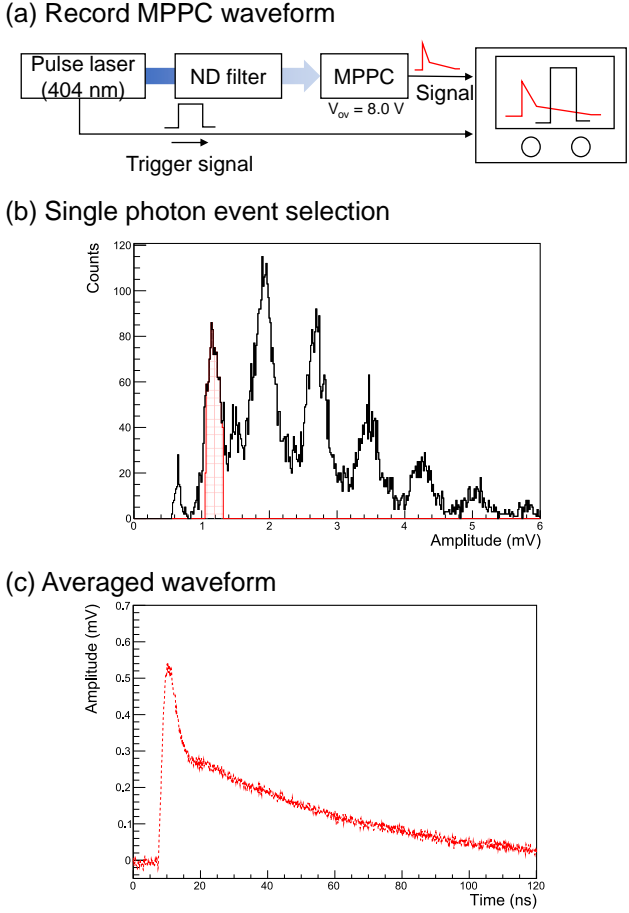

**Fig. 2**. (a) Experimental setup for measuring $(i*s)_{MPPC}(t)$ of the MPPC. Waveforms of the MPPC were recorded using a pulse laser. (b) Histogram of the maximum amplitude was obtained after recording the waveforms; single photon equivalent events (red shaded area) are accepted. (c) The waveforms of the single photon equivalent events are averaged with baseline correction, and the $(i*s)_{MPPC}(t)$ is obtained.

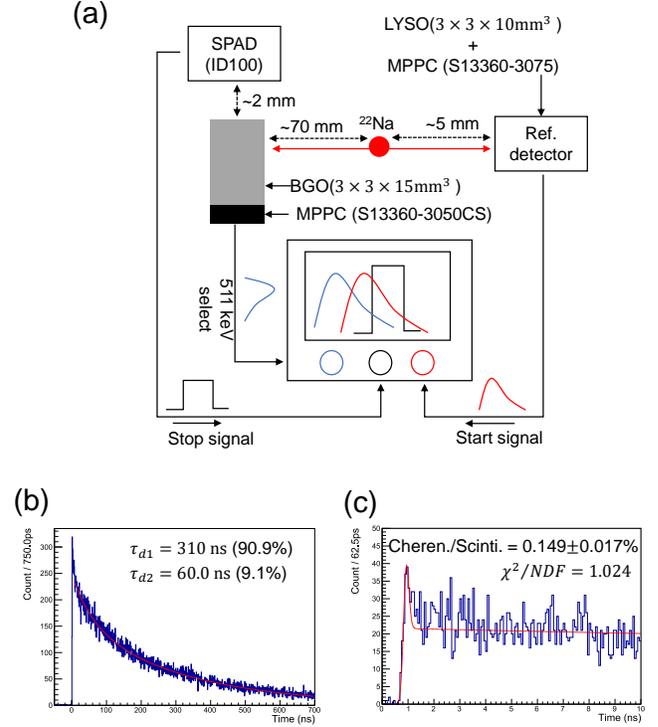

**Fig. 3**. (a) Experimental setup for the TCSPC method to measure the PDF of BGO. The PDF can be measured by the distribution of the time difference between the start and stop signals. (b) BGO scintillation kinetics with the fitting results. (c) The first 10 ns of the BGO scintillation kinetics. At approximately 1 ns, the Cherenkov signal is visible, and its amount corresponds to the result of [12] within an error of 1.5σ. Total measurement time was 27.4 days.

*C. Time correlated single photon counting setup*

To prove that the proposed deconvolution method can retrieve $f_{BGO}(t)$, the conventional TCSPC method was used for the BGO crystal used in this study for comparison. Fig. 3(a) shows the experimental setup for the TCSPC measurement. A reference detector consisting of the $3 \times 3 \times 10$ mm³ LYSO and MPPC (S13360-3075) [24], and SPAD detector (ID100) were used as the start and stop signals, respectively. An additional MPPC (S13360-3050CS) coupled to an unwrapped BGO was used to monitor the energy deposition in the BGO. A $^{22}$Na point source with a radioactive intensity of 2.0 MBq at May 2023 was located between the reference detector and the BGO. With this setup, the coincidence count rate between the three detectors was 0.0299 count per second whereas the coincidence count rate between the reference detector and the additional MPPC was 9.73 count per second (325 times higher), which was calculated without considering the efficiency of the data acquisition system. Assuming that the number of photons entering the SPAD complies with a Poisson distribution, the expected number of photons detected by the SPAD can be calculated as 0.0031 ( log(1+0.0299/9.73)) photons per event. Thus, it was assured that only one photon impinges on the SPAD detector. Figures 3(b) and (c) show the entire and the first 10 ns of the PDF of BGO, respectively, obtained by calculating the time difference between the start and stop signals. The amount of the ratio of the Cherenkov photons to the scintillation photons of 0.149 ± 0.017% corresponds to the result of [11] within an error of 1.5σ. The measurement time was 27.4 days.

*D. Experimental setup for proof-of-concept*

Two experiments were performed to validate the proposed deconvolution method for improved timing performance: recording waveforms from a tested-detector composed of Teflon-wrapped BGO coupled to an MPPC (S13360-3050CS) (1) over several hundred nanoseconds and (2) over the first 10 ns, as illustrated in Figs. 4(a) and (b), respectively.

(1) The objective of the experiment shown in Fig. 4(a) is to validate the accuracy of the proposed deconvolution method by comparing the deconvoluted PDF with that of the TCSPC method. In this experiment, $V_{ov}$ was set to 3.0 V to suppress the unwanted effects of MPPC, that is crosstalk, after pulse, and dark count, which could change the single photon response. The crosstalk probability was less than approximately 5% at $V_{ov}$ = 3.0 V, hence, we considered that these MPPC parameters do not change the single photon response. In addition, a low overvoltage can reduce the PDE of the MPPC, allowing us to make saturation effect negligible; otherwise, it would cause an incorrect PDF. Owing to the simple relationship between the gain and supplied voltage of the MPPC, the single photon response can be easily scaled for this experiment. Specifically, the single photon response obtained in Section *II B* was multiplied by 3.0/8.0. In this study, the worsened SPTR due to a lower overvoltage was not considered for simplicity. The vertical range of the oscilloscope was set at 5 mV/div. for the tested-detector (left detector in Fig. 4 (a)), such that the entire waveform could be accurately measured without vertical saturation of the oscilloscope. The anode signal of the reference detector was directly fed into the oscilloscope with a vertical range of 10 mV/div. to magnify and precisely measure the rising edge of the signal, leading to a reference detector timing resolution of approximately 140 ps FWHM for 511 keV events. The cathode signal of the reference detector was fed into a constant fraction discriminator (CFD), and its signal was used to select a 511 keV event for the reference detector and trigger the oscilloscope. However, the CFD signal was not displayed on the oscilloscope to maintain a sampling rate of 20 GS/s. Otherwise, the sampling rate was reduced to 10 GS/s for the three-channel display. 5,000 waveforms were recorded in this experiment. In the analysis, a histogram of the maximum amplitude of the tested-detector was created and the photoelectric and Compton scattering events were classified. After classification, the waveforms were averaged and deconvolution based on the $(i * s)_{MPPC}(t)$ was performed. The details of the deconvolution method are explained in Section *II E*. The deconvolved PDF was fitted with a function of PDF(t), which is the same as Equation (9) of [25], but slightly modified such that our results could fit more properly. We changed the term representing the prompt emission from

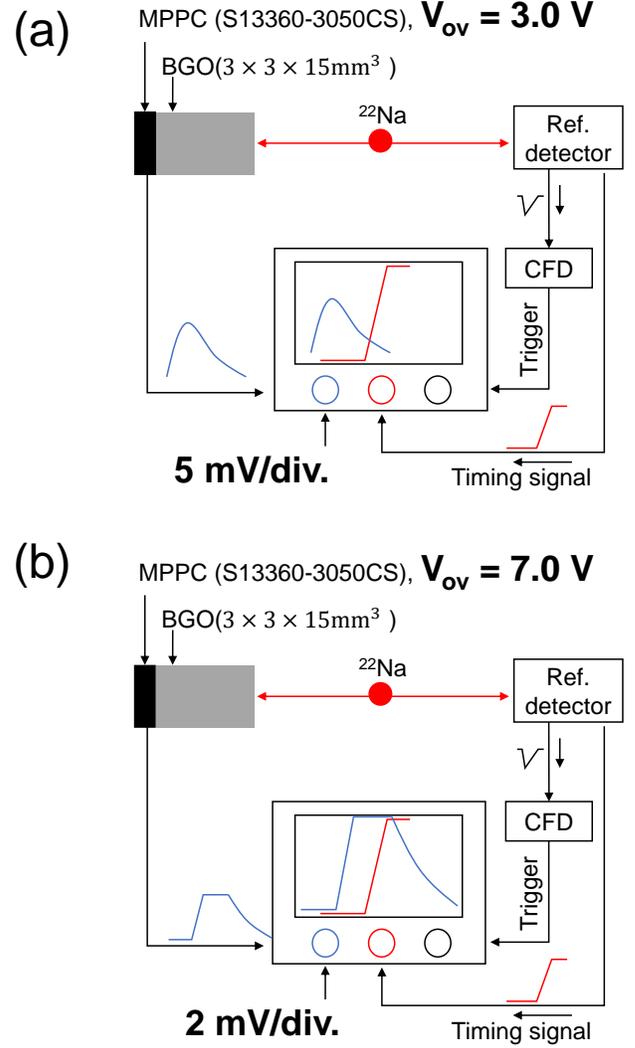

**Fig. 4**. (a) Experimental setup for validating the proposed deconvolution method. The deconvoluted waveforms were compared to the conventional TCSPC method. (b) Experimental setup for investigating a feasibility of extracting Cherenkov photons event-by-event and improving the timing performance.

Gaussian to exponentially modified Gaussian (EMG(t)) as follows:

$$EMG(t) = \frac{A}{2\tau} e^{\frac{1}{2\tau}\left(2\mu + \frac{\sigma^2}{\tau} - 2t\right)} \text{erfc}\left(\frac{\mu + \sigma^2/\tau - t}{\sqrt{2}\sigma}\right), \quad (3)$$

$$\text{erfc}(t) = \int_t^\infty e^{-\lambda^2} d\lambda, \quad (4)$$

where $A, \mu, \sigma$, and $\tau$ are amplitude, mean and standard deviation of the Gaussian, and decay time of exponential, respectively. From the fitting results, the shape of the PDF and the ratio of Cherenkov to scintillation photons were compared with those of the conventional TCSPC method.

(2) The objective of the experiment shown in Fig. 4(b) is to investigate the feasibility of emphasizing the Cherenkov photons and improving the timing performance by deconvolution. $V_{ov}$ was increased to 7.0 V, and the vertical range of the oscilloscope was set to 2 mV/div. such that the rising part of the tested-detector signal could be precisely measured. The setup for the reference detector is the same as that shown in Figure 4(a). In the deconvolution, only the first 12.5 ns of the waveform are cropped to neglect the vertical oscilloscope saturation part of the waveform. In this experiment, 10,000 waveforms were recorded and 6,165 events were accepted as photoelectric events. Event selection was performed using falling edge information from the tested-detector. Although some of the energy information was lost owing to vertical oscilloscope saturation, the falling edge had sufficient information to discriminate Compton scattering events. The deconvolution was performed on an event-by-event basis. The y-axis of the obtained PDF had a dimension of *photons/ps*, and the leading edge threshold was scanned from 0.005 to 0.0385 *photons/50ps* in intervals of 0.02 *photons/50ps* for timing pick-off. For comparison with a conventional leading edge discriminator, the recorded waveforms were directly used, and the voltage threshold was scanned from 0.28 to 3.6 mV in intervals of 0.14 mV for timing pick-off. For both timing pick-off algorithms, the time difference between the reference detector with a fixed timing pick-off threshold was calculated and time difference histograms were generated. The histograms formed an asymmetric shape because of a mixture of Cherenkov- and scintillation-events and were fitted with the exponentially modified Gaussian EMG(t) or a double Gaussian DG(t) defined as follows:

$$DG(t) = \sum_{i=1}^{2} \frac{C_i}{\sqrt{2\pi}\sigma_i} \exp\left(-\frac{1}{2}\left(\frac{x-\mu_i}{\sigma_i}\right)^2\right), \quad (5)$$

where, $C_i$, $\mu_i$, and $\sigma_i$ are the amplitude, mean, and standard deviation of each Gaussian, respectively, depending on their shapes. We selected a fitting function that provides a smaller value of $\chi^2/NDF$ (where NDF denotes the number of degrees of freedom). The errors in the obtained CTRs were calculated systematically by varying the number of bins in the histogram.

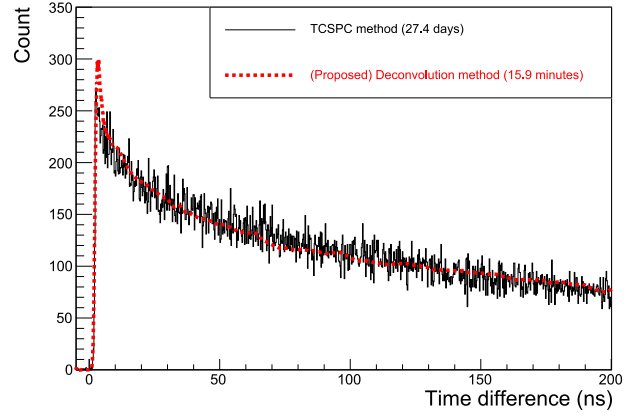

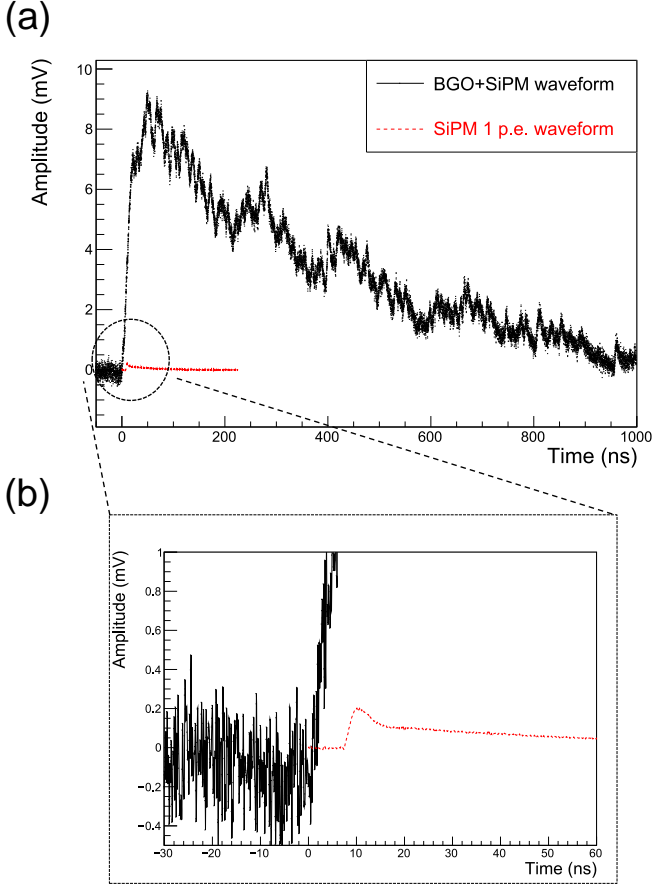

**Fig. 5.** (a) The typical recorded raw waveform from the tested-detector with the superposition of the $(i*s)_{MPPC}(t)$ of the MPPC at $V_{ov}$ = 3.0 V. (b) The magnified waveform and the $(i*s)_{MPPC}(t)$. The amplitude of the single photon is slightly higher than the noise level of the tested-detector.

**Fig. 6.** Deconvoluted and conventional TCSPC-based PDFs. The proposed deconvolution method corresponds to the conventional TCSPC method except for the rising part.

*E. Deconvolution method*

Theoretically, deconvolution can be performed by division of Fourier transforms, particularly in our case, $W(\omega)/(I_{MPPC}(\omega) \times S_{MPPC}(\omega))$. Here, $W(\omega)$, $I_{MPPC}(\omega)$, and $S_{MPPC}(\omega)$ denote the Fourier transforms of $w(t)$, $i_{MPPC}(t)$, and $s_{MPPC}(t)$, respectively. However, once noise was added to the data, this theory did not work, as expected. Therefore, in most cases, deconvolution is performed using iterative methods. The Richardson-Lucy (RL) method is one of the most representative deconvolution algorithms [26,27]. We implemented the RL algorithm to perform the deconvolution of the $(i*s)_{MPPC}(t)$ from the recorded waveform.

As a preprocessing step, the baseline was first corrected to 0 mV for both $(i*s)_{MPPC}(t)$ and the recorded waveforms, and then all data points that were less than 0 mV were set to zero to stabilize the iteration. After correction, for the experiment shown in Fig. 4(a), both the recorded waveforms and $(i*s)_{MPPC}(t)$ were cropped such that the temporal length was 225 ns (4,500 data points) to save computational time. After preprocessing, 200 iterations were performed. The algorithm implemented in this study is described as Algorithm 1.

$$PDF^{(n+1)} = \frac{PDF^{(n)}}{U^T \mathbf{1}} U^T \frac{w}{U \cdot PDF^{(n)}}, \quad (6)$$

where $U$ is an upper triangular matrix consisted of $(i*s)_{MPPC}(t)$.

$$U = \begin{bmatrix} (i*s)_{MPPC}(t_1) & (i*s)_{MPPC}(t_2) & \cdots & (i*s)_{MPPC}(t_N) \\ & (i*s)_{MPPC}(t_1) & \cdots & (i*s)_{MPPC}(t_{N-1}) \\ & & \ddots & \vdots \\ & & \ddots & (i*s)_{MPPC}(t_2) \\ 0 & & & (i*s)_{MPPC}(t_1) \end{bmatrix}$$

where $N$ is the number of data points at 4,500. For the experiment shown in Fig. 4(b), $N_{data\_points}$ and $N_{Iteration}$ were set at 250 (corresponding to 12.5 ns) and 2,000, respectively.

---

**Algorithm 1** Deconvolution of $(i*s)_{MPPC}(t)$ by RL method.

Preprocessing of $w(t)$
$Npe = \frac{\sum_t w(t)}{\sum_t (i*s)_{MPPC}(t)}$
$N_{data\_point} = 4,500$ (or 250)
$N_{Iteration} = 200$ (or 2,000)
$\boldsymbol{weight}(t) = \text{Convolution}(\mathbf{1}, (i*s)_{MPPC}(t))$
$\widehat{\boldsymbol{weight}}(t) = \boldsymbol{weight}(N_{data\_point} - 1 - t)$
$PDF_0(t) = \{ Npe \}/N_{data\_point}$
For n=0 in $N_{Iteration}$
   $\boldsymbol{Forward}_n(t) = \text{Convolution}(PDF_n(t), (i*s)_{MPPC}(t))$
   $\boldsymbol{Error}_n(t) = w(t)/(\boldsymbol{Forward}_n(t) + \varepsilon)$
   $\widehat{\boldsymbol{Error}}_n(t) = \boldsymbol{Error}_n(N_{data\_point} - 1 - t)$
   $\boldsymbol{Backward}_n(t) = \text{Convolution}(\widehat{\boldsymbol{Error}}_n(t), (i*s)_{MPPC}(t))$
   $\boldsymbol{Backward}_n(t) = \boldsymbol{Backward}_n(N_{data\_point} - 1 - t)$
   $PDF_{n+1}(t) = PDF_n(t)\boldsymbol{Backward}_n(t)/(\widehat{\boldsymbol{weight}}(t) + \varepsilon)$
End
$PDF(t) = PDF_{N_{Iteration}}(t)$

---

## III. RESULTS

A typical recorded raw waveform of the tested-detector is illustrated in Fig. 5(a), with the superposition of the $(i*s)_{MPPC}(t)$. An enlarged image is shown in Fig. 5(b). As shown, the amplitude of the $(i*s)_{MPPC}(t)$ was almost the same as the noise level; however, it was slightly higher than the noise level.

The PDFs obtained through the conventional TCSPC and proposed deconvolution methods are overlaid in Fig. 6. The PDF of the deconvolution method was normalized such that the scales of both PDFs matched. The proposed method is consistent with the TCSPC method within a statistical error. Moreover, interestingly, the proposed deconvolution method took only 15.9 min to obtain the PDF, whereas it took 27.4 days in case of the TCSPC method (2,481 times faster). However, the prompt component of the deconvolution method was significantly higher than that of the TCSPC method, which is more clearly visualized and quantified in Fig. 7. Fig. 7(a) shows the amplitude histogram of the tested-detector, which is divided into two classes: photoelectric and Compton scattering events. Figs. 7(b) and (c) show the rising parts of the PDFs of the photoelectric and Compton scattering events, respectively. The best fit results showed that the ratio of prompt to scintillation photons was 0.53% and 0.15% for photoelectric and Compton scattering events, respectively. This significant difference between the two classes led to the conclusion that the prompt component was caused by the Cherenkov photons. Although there is a significant difference between Figs. 3(c) and 7(b), this will be discussed in Section IV (Discussion).

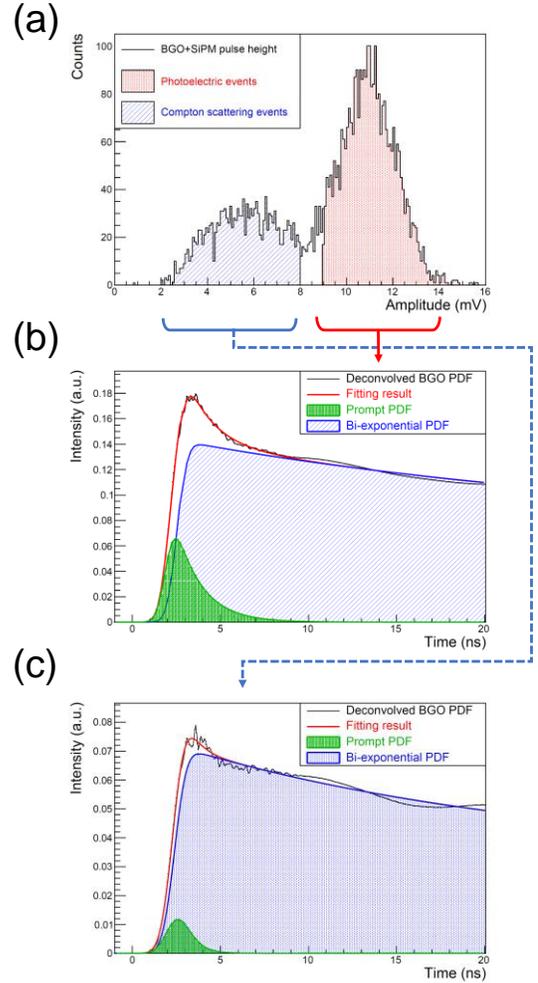

**Fig. 7**. (a) Amplitude histogram of the tested-detector, which is divided into two classes: photoelectric and Compton scattering. (b) The deconvoluted PDF for the photoelectric events. The best fit results show that the amount of prompt component is 0.53% compared to scintillation photons. (c) The deconvoluted PDF for the Compton scattering events. The best fit results show that the amount of prompt component is 0.15% compared to scintillation photons.

Next, we performed deconvolution on an event-by-event basis, according to the experiment shown in Fig. 4(b). Fig. 8(a) depicts not only a raw waveform but also the convolution result of $(i*s)_{MPPC}(t)$ and a deconvoluted PDF (Fig. 8(b)). The convoluted waveform matches the raw waveform, indicating that the deconvolution worked as expected. The trigger timings are different from each other (approximately 6

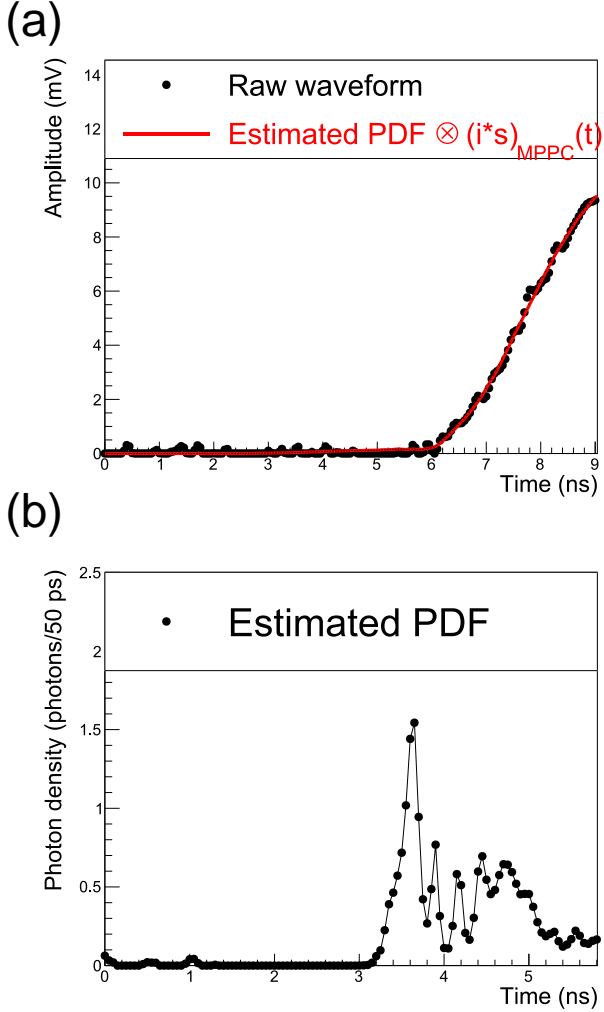

**Fig. 8**. (a) Recorded raw and convolved waveforms using the deconvolved PDF of (b). The convolved waveform corresponds to the raw waveform, indicating that the deconvolution method works as expected.

vs. 3 ns) because $(i*s)_{MPPC}(t)$ has a baseline of ~ 3 ns.

CTRs were evaluated by scanning the leading edge threshold from 0.005 to 0.385 *photons/50ps* and from 0.28 to 3.6 mV for the deconvolved PDF and raw waveform, respectively, as depicted in Fig. 9(a) and (b). The deconvolved PDF clearly exhibits a better CTR than the raw waveform for the entire threshold region. The best CTRs of the deconvolved PDF and the raw waveform were 461.7 ±7.5 and 660.2 ± 9.8 ps FWHM, respectively, as shown in Figs. 10(a) and (b). In addition, the full widths at tenth maximum (FWTMs) were 1296 ±14 and 1493 ± 18 ps, respectively. As explained in Section II *D*, for Fig. 10(a), the function of the double Gaussian DG(t) was used for the fitting procedure owing to the asymmetric shape of the time difference histogram and the smaller $\chi^2/NDF$ than that of the EMG(t) at low threshold levels. The numbers of $\chi^2/NDF$ in Figs. 10(a) and (b) were 1.611 and 1.128, respectively, whereas they were 2.903 and 3.313, respectively, for a single Gaussian fitting.

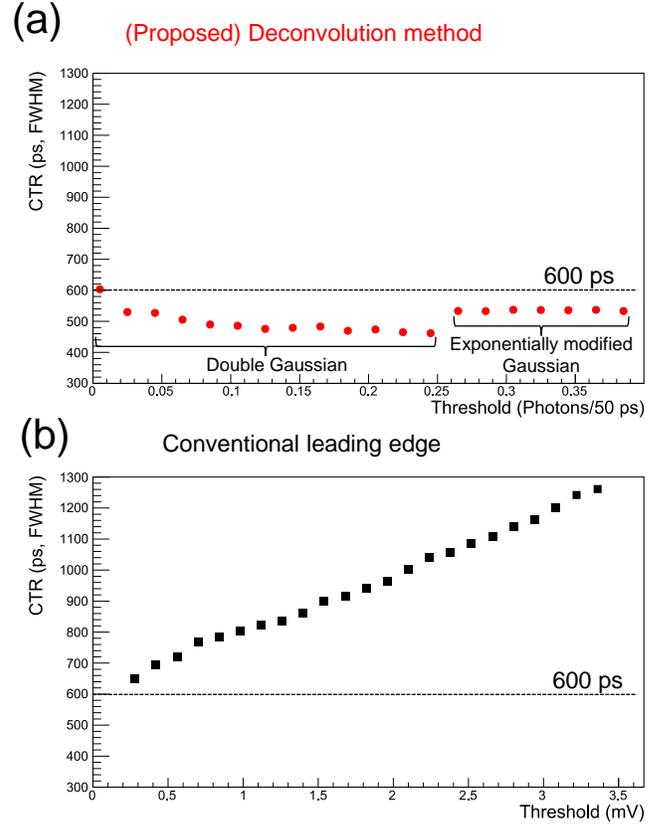

**Fig. 9**. (a) Threshold scan of the deconvolved PDF for CTR calculation. The threshold was scanned from 0.005 to 0.385 *photons/50ps*. Until 0.25 *photons/50ps*, the double Gaussian shows the smaller $\chi^2/NDF$ than that of the exponentially modified Gaussian. (b) Threshold scan of the raw waveform for CTR calculation. The threshold was scanned from 0.28 to 3.6 mV. The deconvolved PDF showed the better CTR than the raw waveform.

## IV. DISCUSSION

A deconvolution method was proposed to emphasize the information of the rising edge of a waveform from a BGO-based detector. The deconvolution method was validated by comparing the deconvolved PDF with the PDF obtained using the conventional TCSPC. Furthermore, the feasibility of improving the timing resolution was validated using event-by-event deconvolution and threshold scans.

The proposed deconvolution method accurately and successfully provided the PDF of BGO, which was consistent with the TCSPC-based PDF, other than the amount of prompt emission. The prompt emissions of the proposed deconvolution and TCSPC methods were 0.53 and 0.149%, respectively. This significant difference was due to the difference between the PDE curve of SPAD (ID100) and MPPC (S13360-3050CS) [11, 28]. Because BGO has a cutoff wavelength of approximately 300 nm, Cherenkov photons with wavelengths down to 300 nm can be transported to the photodetectors. However, SPAD is insensitive to short wavelength photons, whereas the MPPC used in this study can detect them with a PDE of approximately 20% at 300 nm. Equation (1) indicates that shorter wavelength Cherenkov photons are emitted more frequently than those with longer wavelengths. Additionally, the directionality of the Cherenkov emission in BGO might have caused the significant difference. According to [29, 30], the first several Cherenkov photons were emitted in the forward direction. Hence, the Cherenkov photons partially escaped from the BGO in the case of the TCSPC setup because BGO was unwrapped and irradiated with gamma rays perpendicular to the long side of the crystal, causing a reduction in the prompt component. Conversely, in the experimental setup shown in Fig. 4(a), because BGO is wrapped with Teflon, all the Cherenkov photons emitted can be detected by the MPPC without escaping from the crystal.

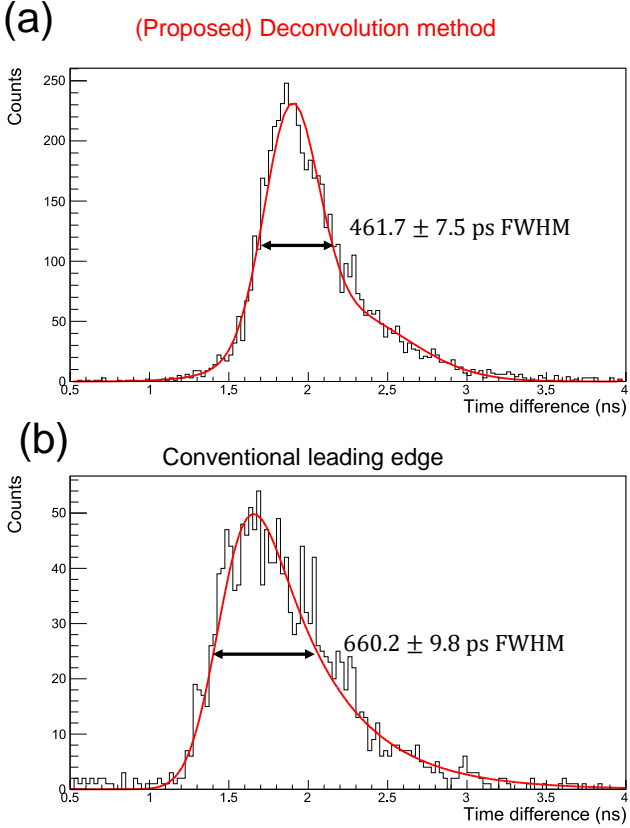

**Fig. 10**. (a) Best CTR of the deconvolved PDF fitted by double Gaussian $DG(t)$. $\chi^2/NDF$ was 1.611, while that of the $EMG(t)$ was 1.666. (b) The best CTR of the raw waveform fitted by $EMG(t)$. $\chi^2/NDF$ was 1.128. Both histograms showed the asymmetric shapes which are caused by the mixture of Cherenkov-triggered and scintillation-triggered events. In addition, a parameter of $\tau$ which can be considered slow scintillation events shows a significant difference between the deconvolved PDF and the raw waveform.

The prompt component of the deconvolved PDF is represented by EMG(t) rather than a Gaussian. This may have been affected by the number of iterations (200). The value of 200 may be slightly insufficient to completely recover the prompt component because it contains a high frequency response. Although an increased number of iterations can recover high frequency components, it will simultaneously increase the high frequency noise, as discussed in [23]. In fact, the distribution of the prompt components drastically fluctuated and lost quantitativity when the number of iterations was increased to more than 300. Furthermore, the shape of the SPTR of the MPPC would also affect the shape of the prompt component obtained in this study because the shape of the SPTR of an SiPM is generally not Gaussian but an EMG as reported in [31]. Apart from that, the prompt component shows a much wider distribution than that shown in Fig. 3(c), although it should be as fast as an SPTR of approximately 150 ps FWHM [11]. In the experiment shown in Fig. 4(a), the vertical range of the oscilloscope is set at 5 mV/div., and $V_{ov}$ was decreased to 3.0 V; thus, $(i*s)_{MPPC}(t)$ was almost buried in the electric noise, making the SPTR on the order of nanosecond according to the equation below [9].

$$\sigma_t = \frac{\sigma_{noise}}{\left(\frac{dV}{dt}\right)}, \qquad (7)$$

where $\sigma_t$, $\sigma_{noise}$, and $dV/dt$ are the timing uncertainty, electric noise, and slew rate of the single photon response, respectively. Furthermore, $(i*s)_{MPPC}(t)$ used for the deconvolution contained information on the SPTR at $V_{ov}$ = 8.0, rather than at $V_{ov}$ = 3.0 V, for simplicity. Therefore, a worse SPTR should be considered to reproduce the prompt component more precisely. The prompt part of the deconvolved PDF could be as sharp as that of the TCSPC if $V_{ov}$ could be increased while suppressing unwanted SiPM parameters such as crosstalk, because the SPTR can be improved according to $V_{ov}$.

It should be noted that the $V_{ov}$ of the MPPC should be sufficiently low to avoid unwanted crosstalk, after pulse, and saturation effects. As a preliminary experiment, a deconvolved PDF of an LYSO with dimensions of 3 × 3 × 10 mm³ coupled to the MPPC (S13360-3050CS) was also compared with the TCSPC method, and the results showed good agreement as long as the saturation effect of the MPPC was negligible. However, when the MPPC of S13360-3075CS was used as a photodetector, the MPPC saturation effect rapidly became explicit, causing a PDF inconsistent with that of the TCSPC.

$(i*s)_{MPPC}(t)$ has a long tail of 50–100 ns because of its electric features. However, the single photon response of a PMT or microchannel plate PMT is as fast as or faster than 1 ns. Therefore, the temporal shape of the single photon response is minor compared with a $f_{BGO}(t)$, and a PDF could be obtained without a deconvolution process [32,33]. However, if the decay time constants of the scintillators are faster than the single photon response of the PMT, the deconvolution method is useful even for PMT-based detectors. For example, barium fluoride, which is an interesting scintillator for fast-timing applications, has recently received attention [34,35] and may benefit from the deconvolution method for its evaluation. Fast perovskite scintillators would also benefit from this process [36].

The event-by-event deconvolution demonstrated a 43 and 15% improvement in the CTR in FWHM and FWTM, respectively, compared with the conventional voltage-based leading edge discriminator, as depicted in Figs. 9 and 10, respectively. This can be concluded as a proof of emphasized detection of Cherenkov photons because the two Cherenkov- and scintillation-triggered events were clearly separated as $\chi^2/NDF$ of DG(t) was smaller than that of EMG(t) at a threshold level lower than 0.25 *photons/50ps*. A high threshold level loses the chance to trigger Cherenkov photons, and Cherenkov- and scintillation-triggered events gradually become invisible. Thus, the histogram is shaped like an EMG. At *0.385 photons/50ps*, $\chi^2/NDF$s of DG(t) and EMG(t) were 1.78 and 1.20, respectively. The asymmetric shape of the histogram represented by EMG(t) is the nature of the fact that two different types of detectors are used for the evaluation, as reported in [37]. The moderate improvement in the FWTM, compared to the FWHM was due to the remaining scintillation-triggered events, which accounted for 42.6% of the total events.

In this study, no fast amplifiers that can improve the intrinsic SPTR, such as those in [9,10,38,39], were used. Therefore, the absolute number of CTRs obtained was still worse than that of state-of-the-art BGO-based detectors [19, 40]. Although it is expected that using fast amplifiers would improve CTRs, electric saturation should be carefully considered, particularly for high-gain amplifiers, because electric saturation would cause incorrect PDFs.

The numbers of counts displayed in Figs. 10(a) and (b) are 5,755 and 1,633 events, respectively. This large difference can be explained by the robustness of each timing pick-off algorithm to electrical noise. The conventional voltage-based leading edge discriminator easily suffers from electric noise at low threshold levels and consequently picks up incorrect timing information, although its CTR might potentially be better for thresholds lower than 0.28 mV. To obtain the same number of events as 5,755, the threshold level should be increased; however, this significantly affects the CTR. However, the proposed deconvolution method is more tolerant to noise, enabling us to determine the optimal threshold level.

In addition to emphasizing the Cherenkov photon signal, the proposed deconvolution method has interesting potential. The deconvolution method can recover the timing information of all the detected photons over the deconvolution range. This indicates that even the timing information points of 100 photons can be used to estimate the gamma-ray interaction time. Once multiple timing information is available, the Cramer-Rao lower bound can be reached, and an improved CTR is expected because the Cramer-Rao lower bound is the theoretical limit of the best CTR [41-43]. Furthermore, time-based energy discrimination and/or depth-of-interaction estimation would be considerable with plural timing information, which is indispensable for achieving an ultimate CTR of 10 ps FWHM [44. 45].

The most important limitation of this study, which should be eagerly addressed, is that the proposed deconvolution method requires waveforms. Even for the first 3–4 ns, the waveform contained approximately 100 data points (Fig. 8(a)). This indicates that the current research is still in the laboratory stage. Therefore, as the next step, we will focus on making the proposed method more practical such that it can be implemented in a system. Reducing the number of data points to less than 10 has great potential for implementation in a system or a field programmable gate array.

## V. CONCLUSION

In this study, we proposed a method in which a single photon response was deconvolved from a BGO-based detector waveform to emphasize the Cherenkov photon signal and improve the timing resolution of the detector. As a proof of concept, the deconvolution method was validated by obtaining a PDF of BGO and comparing it with that of a conventional TCSPC method. The results of the proposed deconvolution method were in good agreement with those of the TCSPC method. In addition, the CTR was improved using the proposed deconvolution method. This can be concluded as proof of the emphasized Cherenkov photon detection. The proposed deconvolution method has the potential to reach the Cramer-Rao lower bound on the timing resolution and room for further improvement in timing performance when combined with MPPCs and fast amplifiers.

ACKNOWLEDGMENT

The authors acknowledge Yuya Onishi from Hamamatsu Photonics K.K. for insightful discussions.